\documentclass[journal=langd5,manuscript=article]{achemso}

\usepackage{graphicx}
\usepackage{dcolumn}
\usepackage{bm}
\usepackage{epsfig}
\usepackage{amssymb}
\usepackage{psfrag}
\usepackage{natbib}
\usepackage{epstopdf}
\usepackage{color}
\usepackage[version=3]{mhchem} 
\usepackage[T1]{fontenc}       

\newcommand{\beq}{\begin{equation}}
\newcommand{\eeq}{\end{equation}}
\newcommand{\beqa}{\begin{eqnarray}}
\newcommand{\eeqa}{\end{eqnarray}}
\newcommand{\p}{\partial}

\author{Sandip Ghosal}
\affiliation{Department of Mechanical Engineering \& 
Engineering Sciences and Applied Mathematics, 
Northwestern University, Evanston, IL 60208, USA}
\email{s-ghosal@u.northwestern.edu}
\author{John D. Sherwood}
\affiliation{Department of Applied Mathematics and Theoretical Physics, University of Cambridge, Cambridge CB3 0WA, UK}

\title[Repulsion Between Finite Charged Plates]{Repulsion Between Finite Charged Plates with Strongly Overlapped
Electric Double Layers}

\begin{document}
\begin{abstract}
The screened Coulomb interaction between uniformly charged flat plates is considered at very small plate separations for which the Debye layers are strongly overlapped, in the limit of small electrical potentials. If the plates are of infinite length, the disjoining pressure between the plates decays as an inverse power of the plate separation. If the plates are of finite length, we show that screening Debye layer charges close to the edge of the plates are  no longer constrained to stay between the plates, but instead spill out into the surrounding electrolyte.
The resulting reduction in the disjoining pressure is calculated analytically. 
A similar reduction of disjoining pressure due to loss of lateral confinement of the Debye 
layer charges should occur whenever the sizes of the interacting charged objects 
become small enough 
to approach the Debye scale. We investigate the effect here in the context of a 
two dimensional model problem that is sufficiently simple to yield analytical results.
\end{abstract}

The theory of the screened Coulomb DLVO repulsive force between 
colloidal particles with like charge, due to Derjaguin and 
Landau~\cite{DL41}
and Verwey and Overbeek~\cite{verwey_1948},  underpins our understanding of the stability of colloidal suspensions against 
flocculation and has subsequently been improved and extended~\cite{mccartney_improvement_1969,bell_approximate_1970,white_deryaguin_1983,bhattacharjee_surface_1997,schnitzer_generalized_2015}. Careful measurements using atomically smooth mica surfaces \cite{israelachvili_measurement_1978,israelachvili_solvation_1987}, atomic force microscopes \cite{todd_probing_2004} and laser optical tweezers \cite{sugimoto_direct_1997,gutsche_forces_2007} have confirmed the theory within its expected range of validity. 
Increased interest in non-DLVO repulsive interactions
acting at separations of only a few nanometers, a distance of the
same order as that of the attractive 
dispersion forces~\cite{israelachvili_international_1977,ninham_progress_1999},
necessitates
a careful analysis of DLVO forces at very 
small particle separations where the classical weak overlap
approximation for the electric potential between the particles fails. 
Philipse {\it et al.} proposed an 
alternative approach~\cite{philipse_algebraic_2013} in which the 
double layers are so strongly overlapped that the space between the 
two opposing surfaces is essentially at uniform potential. 
For surfaces of fixed charge, the resulting disjoining pressure was shown~\cite{philipse_algebraic_2013}  to 
diverge algebraically with the gap width. 
Here we revisit this regime of strongly overlapped Debye layers
but suppose that the plates confining the electrolyte are of finite
lateral extent. This may be regarded as the simplest instance of a broader
class of problems where there is a loss of confinement of the Debye layer on a
lateral scale that is comparable in magnitude to the Debye length.
Examples include plate like clay particles \cite{olphen_introduction_1963,callaghan_interparticle_1974,
leote_de_carvalho_nonlinear_2000}, nanocolloids and the tips of AFMs.

We show that in this case some of the neutralizing charge between 
the opposing faces spills out of the gap at the edge of the plates, resulting in a 
reduction of the disjoining pressure. Such loss of confinement of the
Debye layer charge has been previously investigated in
cylindrical nanopores~\cite{sherwood_electroosmosis_2014}
in the context of electroosmosis, and
leads to a significant reduction of electroosmotic flux.
It can be seen in numerical computations\cite{leote_de_carvalho_nonlinear_2000}, but its
effect upon the force between particles has not been investigated analytically.

\begin{figure}[t]  
  \centering
  \includegraphics[width=0.8\textwidth]{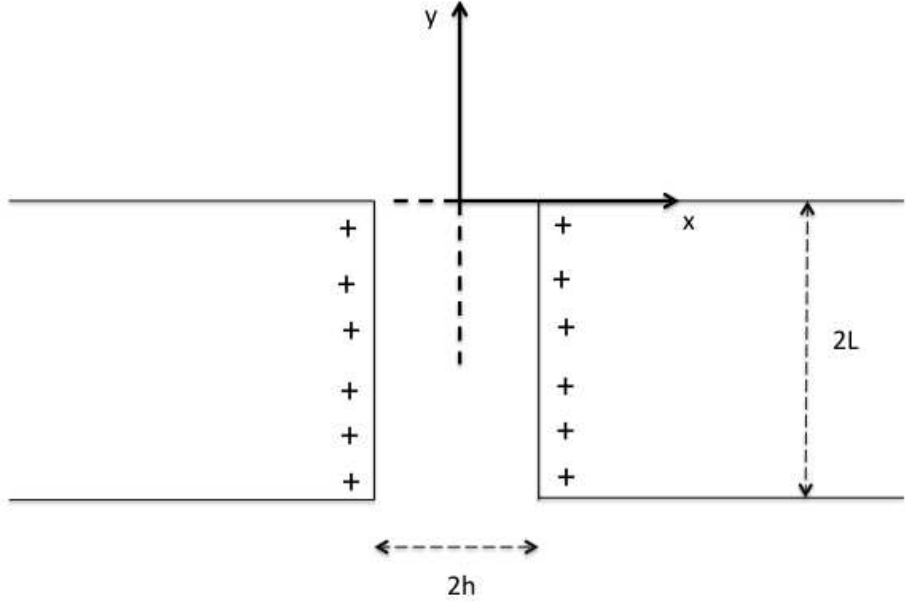}
      \caption{Sketch showing the geometry of the problem and the system of co-ordinates.}
      \label{fig_geom}
\end{figure}
We consider the simplest possible plane 2D model:
a dielectric slab of thickness $2L$ split into two parts by a slot of
width $2h\ll L$ occupying the region $-h<x<h$, as shown
in Figure~\ref{fig_geom}.  The region
outside the solid is occupied by an electrolyte (permittivity
$\epsilon$, Debye length $\kappa^{-1}\gg h$). There is fixed surface charge
of density $\sigma$ on each of the opposing faces $x=\pm h$, but the rest
of the interface, on $y=0$ and $y=-2L$, is uncharged. 
We assume 
that the electric potential $\phi$ is small, with $e\phi/(kT)\ll 1$,
where $e$ is the elementary charge and $kT$ the Boltzmann temperature, so that the Debye-H\"{u}ckel (DH) limit~\cite{russel_saville_schowalter} is appropriate.
We assume that $L,\kappa^{-1} \gg h$, but 
 $\kappa L$ is arbitrary, that is, the length of the channel 
may be large or small compared to the Debye length. 
When $\kappa L \rightarrow \infty$ 
our problem is identical to that considered by Philipse et al.\cite{philipse_algebraic_2013}
except that we restrict our attention to the low potentials appropriate 
for the DH approximation.
This restriction requires that the surface
charge density $\sigma$ on the surface of the plates should be
small, with $\sigma\ll \epsilon\kappa^2hkT/e$, but leads to simplifications that
enable us to investigate the overspill of ions analytically.

We first determine the equilibrium potential $\phi (x,y)$ by solving the DH
equation~\cite{russel_saville_schowalter} $\nabla^2 \phi = \kappa^2 \phi$ in the fluid domain 
with Neumann boundary conditions:
$-\epsilon \p_{n} \phi$ equals $\sigma$ on the opposing faces ($x=\pm h$, $-2L<y<0$) 
but is zero elsewhere on the boundary.
Thus, we assume that the ratio of the permittivity of the solid to that of the fluid is sufficiently small 
that there is no incursion of field lines into the solid. This commonly 
invoked approximation is reasonable 
because the dielectric constant of water ($\sim 80$) is much larger than the corresponding  
values ($\sim 2$--$4$) for most nonpolar solid substrates. If the channel is infinitely long
the potential
between the faces is $\phi=(\sigma/\epsilon\kappa)\cosh(\kappa x)/\sinh(\kappa h)
\approx\phi_m\cosh(\kappa x)$, where $\phi_m=\sigma/(\epsilon\kappa^2h)$.
The neutrality length $\lambda$ defined by Philipse et al.\cite{philipse_algebraic_2013}
is therefore $\lambda=2 he\phi_m/(kT)$ and the DH approximation corresponds to $\lambda\ll h$.

More generally, when $L$ is finite, we define an average potential 
$\overline{\phi} (y) = (2h)^{-1} \int_{-h}^{+h} \phi (x,y) \, dx$
across the channel. 
Integrating the DH equation across the channel and using the Neumann 
boundary conditions on the walls, we derive 
\begin{equation} 
\frac{d^2 \overline{\phi}}{d y^2} = \kappa^{2} \overline{\phi} - \frac{\sigma}{\epsilon h}.
\end{equation} 
Since $\kappa^{-1} \gg h$ the equipotential surfaces are almost orthogonal to the channel walls. Thus, 
\begin{equation} 
\phi(x,y) \sim \overline{\phi}(y) = \frac{\sigma}{\epsilon \kappa^2 h} + A \cosh [\kappa (y+L)],
\label{phi_within0}
\end{equation} 
with error $O(\kappa h)^{2}$.

Next we consider the behavior of $\phi$ outside the channel 
at distances $r = \sqrt{x^2+y^2} \gg h$. In this outer region the potential has 
cylindrical symmetry and must vanish at infinity. Therefore 
\begin{equation} 
\phi (x,y) = B K_{0} (\kappa r),
\label{K0} 
\end{equation} 
where $K_0$ is the modified Bessel function of the second kind.
In order to determine $A$ and $B$ we desire that $\phi$ and its derivative match smoothly to the interior potential, Eq.~(\ref{phi_within0}),
at the channel entrance. This is, however, impossible as 
 $K_{0} (\kappa r)$ diverges when $r \rightarrow 0$ and the potential Eq.~(\ref{K0}) ceases to be valid when 
$r \lesssim h$. To circumvent this difficulty we consider an intermediate zone
consisting of a region of size $\sim h$ outside the pore entrance in order to ``bridge'' the two potentials.
 
In the intermediate zone, the potential changes over distances of order $h$. Thus, $\nabla^{2} \phi \sim \phi / h^2 \gg \kappa^2 \phi$, so that the DH equation may be approximated by Laplace's equation in this region. Specifically, we need a solution of the Laplace equation consistent with the mixed boundary condition 
\begin{eqnarray} 
\phi(x,0) = \frac{\sigma}{\epsilon \kappa^2 h} + A  \cosh ( \kappa L) \quad  && \text{if} \: |x| \leq h, \label{phibc0} \\
\partial_{y} \phi(x,0) = 0  \quad && \text{if} \: |x| > h,
\label{phibc}
\end{eqnarray} 
and having an electric flux 
\begin{equation} 
F_e = - \int_{-h}^{h} \frac{\partial \phi}{\partial y} (x,0) \; dx = - 2 \kappa h A \sinh ( \kappa L).
\label{flux}
\end{equation} 
We note that since the field vanishes at infinity, the flux $F_e$ is simply related to the charge $\Delta Q$ that
spills out of the channel at $y=0$:
\begin{equation} 
 \Delta Q = - \epsilon F_e = 2 \epsilon \kappa h A \sinh ( \kappa L).
 \label{LostCharge}
 \end{equation}
 A harmonic potential satisfying boundary conditions Eqs.~(\ref{phibc0})--(\ref{flux}) 
 may be found by considering the solution
 $\Phi$ for potential flow through a slit~\cite{milne-thomson_theoretical_1968}, given by
 \begin{equation} 
 \frac{x^2}{h^{2} \cosh^{2} \Phi} + \frac{y^2}{h^{2} \sinh^{2} \Phi} = 1.
 \label{MT}
 \end{equation} 
 The equipotential lines are ellipses and $\Phi$ satisfies the boundary conditions
 \begin{eqnarray} 
\Phi(x,0) = 0 \quad  && \text{if} \: |x| \leq h, \\
\partial_{y} \Phi(x,0) = 0  \quad && \text{if}  \: |x| > h.
\end{eqnarray} 
To determine the flux, we consider Eq.~(\ref{MT}) for $|x|<h$ in the limit
$y \rightarrow 0$, for which
$\Phi \rightarrow 0$, $\sinh \Phi \approx \Phi$ and
$\cosh \Phi \approx 1$.
 Eq.~(\ref{MT}) then reduces to 
\begin{equation}
\Phi(x,y) \approx \frac{y}{\sqrt{h^{2} - x^{2}}}
\end{equation} 
in the region $y > 0$ and $|x| < h$. Therefore, the flux is 
\begin{equation} 
- \int_{-h}^{+h} \frac{\partial \Phi}{\partial y} \; dx = - \int_{-h}^{+h} \frac{dx}{\sqrt{h^{2} - x^{2}}} 
 = - \pi.
 \end{equation} 
The solution satisfying boundary conditions Eqs.~(\ref{phibc0})--(\ref{flux}) can therefore be expressed as 
 \begin{equation} 
 \phi (x,y) =A \cosh( \kappa L) + \frac{\sigma}{\epsilon \kappa^2 h}    +  \frac{2}{\pi}  \kappa h A \Phi   \sinh(\kappa L).
 \label{sol4phi}
 \end{equation} 
 
We now match the solution in the far field, Eq.~(\ref{K0}),
to that in the intermediate zone, Eq.~(\ref{sol4phi}).
The asymptotic properties\cite{abramowitz}  of 
the Bessel function $K_0$ for small $\kappa r$
imply that the inner expansion of the outer potential
(\ref{K0}) is
\begin{equation} 
 \phi (x,y) = B K_{0} (\kappa r) \sim - B \left[    \gamma + \ln \left( \frac{\kappa}{2} \right) \right] - B \ln r,
 \label{innerlimit} 
 \end{equation} 
where $\gamma \approx 0.577$ is Euler's constant.
 Next, we consider the asymptotic form of the intermediate solution (\ref{sol4phi}) at large distances from the pore. In this region $\Phi$ 
 is large, and $\cosh \Phi \sim \sinh \Phi \sim \exp(\Phi) /2$
 so that, by Eq.~(\ref{MT}), $\Phi \sim \ln (2r/h)$. Thus the outer
expansion of the intermediate solution (\ref{sol4phi}) takes the form 
\beqa
\phi &\sim&  \frac{\sigma}{\epsilon \kappa^2 h} + A \left[ \cosh(\kappa L) - \frac{2}{\pi} 
 \kappa h \sinh( \kappa L)  \ln \left( \frac{h}{2} \right) \right] \nonumber \\
&+& \frac{2}{\pi}  \kappa h A \sinh (\kappa L)  \ln r .
\label{outerlimit} 
\eeqa
Equating the coefficients in Eqs~(\ref{innerlimit}) and (\ref{outerlimit}) we obtain
\beqa
A &=& - \frac{\kappa \Lambda}{\sinh (\kappa L)}  \phi_m, \label{eq4A} \\
B &=& - \frac{2}{\pi}  \kappa h A \sinh(\kappa L), \label{eq4B}
\eeqa
where the length $\Lambda$ is defined by 
\begin{equation} 
(\kappa \Lambda )^{-1} = \coth( \kappa L) + \frac{2}{\pi} \kappa h 
\left\{ \ln \left( \frac{4}{\kappa h} \right) - \gamma \right\}
\label{lostL}
\end{equation}
and
$\phi_m = \sigma / (\epsilon \kappa^2 h)$
is the potential between infinite plates. 

\begin{figure}[t]   
  \centering
  \includegraphics[width=0.8\textwidth]{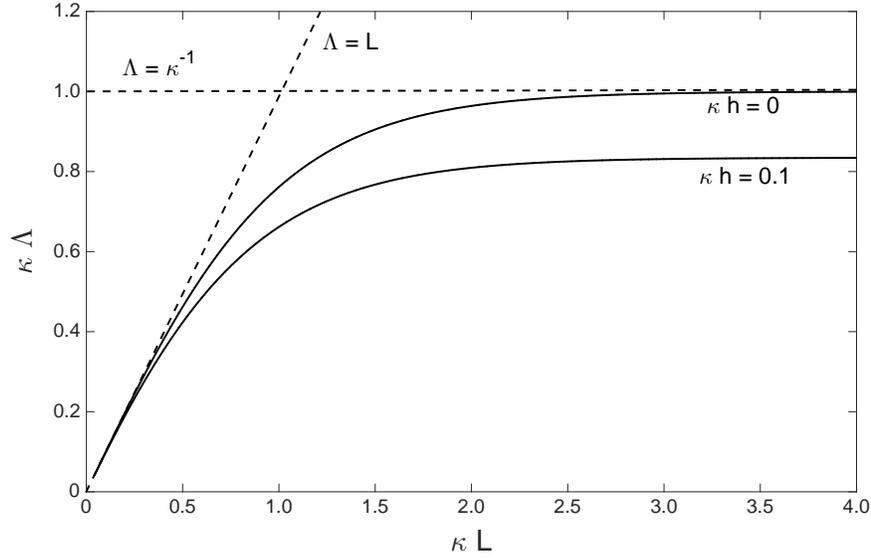}
      \caption{Normalized ``lost length'' $\kappa \Lambda$ [Eq.~(\ref{lostL})] as a function 
      of $\kappa L$ for fixed values of $\kappa h$.}
      \label{fig_LostLen}
\end{figure}
The length $\Lambda$ has a simple physical interpretation. 
Combining Eqs~(\ref{LostCharge}) and (\ref{eq4A}) we see that 
$ \Delta Q = - 2 \sigma \Lambda$. Thus, the neutralizing charge that would normally be confined 
within a length $\Lambda$ of the channel at either end now spills out into the surrounding 
electrolyte.
From Eq.~(\ref{lostL}) we see that, to leading order in $\kappa h$, 
this ``lost length'' $\Lambda \sim \kappa^{-1} \tanh \kappa L$. Thus, $\Lambda$
increases from 0 to $L$ 
as the Debye length $\kappa^{-1}$ increases from zero to infinity.
The dependence of $\kappa \Lambda$ on $\kappa L$ for several fixed values of $\kappa h$
 is shown in Figure~\ref{fig_LostLen}.

\begin{figure}[t]   
  \centering
    \includegraphics[width=0.8\textwidth]{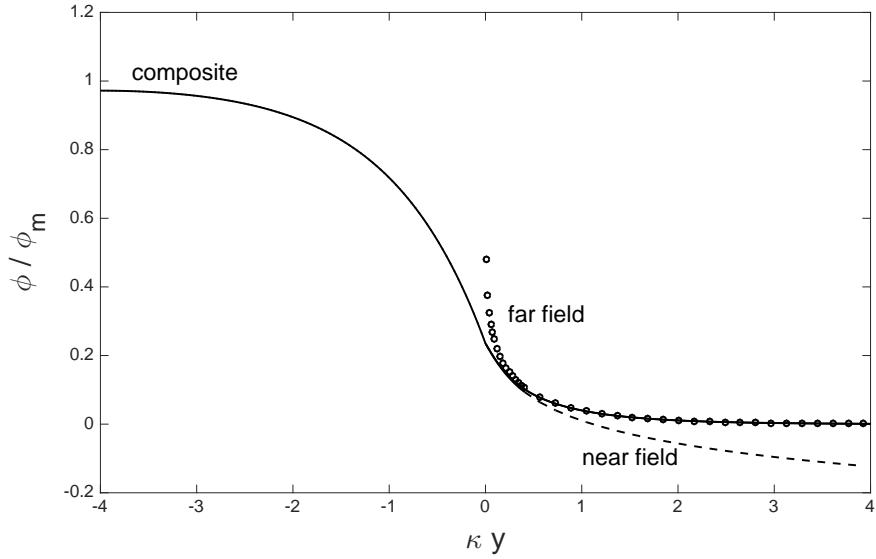}
      \caption{Normalized potential $\phi / \phi_m$ 
on the symmetry axis $x=0$, plotted against $\kappa y$. The solid line is the composite solution given by Eq.~(\ref{phi_within0}) in $y \leq 0$ and Eq.~(\ref{composite}) in $ y > 0$.
In $y >0$, the intermediate field Eq.~(\ref{sol4phi}) [dashed line] and far field Eq.~(\ref{K0}) [symbols], are 
also shown. Parameter values are $\kappa h = 0.1$ and $\kappa L = 4$ which give [Eq.~(\ref{lostL})]
$\kappa \Lambda = 0.764$. }
 \label{fig_comp}
\end{figure}

We can now construct a composite solution~\cite{Hinch}, uniformly valid in $y>0$, by adding Eqs (\ref{K0}) 
 and (\ref{sol4phi}) and subtracting the solution in the overlap region, Eq.~(\ref{innerlimit}). 
 We obtain, after simplification using Eqs (\ref{eq4A}) and (\ref{eq4B}),
 \begin{equation} 
 \phi = \frac{2}{\pi} \kappa^2 h \Lambda \phi_m \left[ \ln \left( \frac{2r}{h} \right) + K_0 (\kappa r) - \Phi \right].
 \label{composite}
 \end{equation} 
 An example of the intermediate [Eq~(\ref{sol4phi})], far field [Eq.~(\ref{K0})]
and composite [Eq.~(\ref{composite})]  potentials for 
 fixed $\kappa h$ and $\kappa L$ is shown in Figure~\ref{fig_comp},
where the potential $\phi(0,y)$ along the symmetry axis $x=0$ is plotted.
 Also shown is the solution in the interior of the channel
 [Eq.~(\ref{phi_within0})]. Note that there is a discontinuity in
 $\partial\phi/\partial y$ at $x=y=0$, though continuity of the mean
value of $\partial\phi/\partial y$ across the gap between the plates at
the edge $x=0$ is ensured by Eq. (\ref{flux}).
 The discontinuity could be eliminated by using
 the solution of the Laplace equation~\cite{yariv_application_2015} that holds both outside and
 inside the channel and matches onto both the
 far-field Eq.~(\ref{K0}), and interior  Eq.~(\ref{phi_within0}) solutions, rather than simply 
 `patching' the intermediate zone to the solution inside the
channel. However, the results of such matching differ little from
those obtained here that merely ensure continuity of $\phi$ and of the total flux $F_e$ 
at $y=0$.

The repulsive force on either plate is given~\cite{russel_saville_schowalter} by the integral
\beq
F = \int_{-L}^{\infty} dy \,\left[ \epsilon  \kappa^2 \phi^2 + \epsilon \left( \frac{\p \phi}{\p y} \right)^{2} 
\right]_{x=0}.
\label{gen_force}
\eeq 
It is convenient to express our results in terms of the pressure scale $f_{0} = \sigma^2 / (2 \epsilon)$.
Using Eq.~(\ref{phi_within0}), we find the contribution $F_1$ to the force integral Eq.~(\ref{gen_force}) 
from the region $-L<y<0$ within the channel:
\beq 
\frac{F_1}{2L} = \frac{f_0}{\kappa^2 h^2} \left[ 
1 - \frac{2 \Lambda}{L}  +
\frac{\kappa \Lambda^2}{L} \coth (\kappa L) 
\right]. 
\label{eq4F1}
\eeq 

The contribution from the part of the force integral in Eq.~(\ref{gen_force}) that lies outside the channel 
consists of the contribution $F_2$ from the osmotic term that is proportional to $\phi^2$ and 
the electric stress contribution $F_3$ that is proportional to $(\p_y \phi)^2$. 
After integration, these contributions are respectively 
$F_{2} = 8 \kappa \Lambda^2 f_0  G (\kappa h) / \pi^2$ and 
$F_{3} = 8 \kappa  \Lambda^2 f_0  H (\kappa h) / \pi^2$
 where the functions $G(x)$  and $H(x)$ are defined as 
 \beqa 
 G(x) &=&  \int_{0}^{\infty}dt  \left[  K_{0} (t) + \ln \left\{ \frac{2t}{
t + \sqrt{t^2 + x^2} }  \right\} \right]^{2}, \\
  H(x) &=& \int_{0}^{\infty} dt \left[ \frac{1}{t} - \frac{1}{\sqrt{x^2 + t^2}} - K_1 (t) \right]^2.
 \eeqa 
These integrals are well defined, and~\cite{kolbig_two_1995} when $x\ll 1$
 \beqa 
 G(x) &\sim& G(0) = \int_{0}^{\infty} K_{0}^{2} (t) \; dt = \frac{\pi^2}{4},
 \label{Gzero} \\
  H(x) &\sim& \int_{0}^{\infty} \frac{dt}{x^2 + t^2} = \frac{\pi}{2 x}.
 \eeqa
 Thus, for small values of $\kappa h$, we have, $F_2 = 2 \kappa \Lambda^2 f_0$ and 
 $F_{3} =  4 \Lambda^2 f_0 / (\pi h)$. 
 The total force is $F = F_1 + F_2 + F_3$ and the disjoining pressure $f = F / (2L)$ is
\beq
\kappa^{2}  h^{2} \frac{f}{f_0} = 
1 - \frac{2 \Lambda}{L} 
+ \frac{\kappa \Lambda^2}{L} 
 \left\{ \coth(\kappa L) + \frac{2 \kappa h}{\pi} + \kappa^2 h^{2} \right\} .
 \label{totalF}
\eeq 
If only the leading order in $\kappa h$ is retained, we have the simpler result
\beq
\frac{f}{f_0} =  \frac{1}{\kappa^{2} h^{2}} \left[ 
 1 - \frac{\tanh (\kappa L) }{\kappa L} \right].
\label{totalF1}
\eeq
In the limit $\kappa L \rightarrow 0$,
when $h\ll L\ll \kappa^{-1}$,
Eq.~(\ref{totalF1}) shows that 
the average disjoining pressure
$f \sim (f_0/3) (L / h)^{2}$ is independent of the Debye length.

The average disjoining pressure in the limit
$\kappa L \rightarrow 0$
can also be derived by considering the free energy of the system.
When $L \gg h$, the 
electric field $E(y)$ is predominantly in the $y$ direction and must satisfy the
condition of flux conservation: $2 h \epsilon E(y) = 2 \sigma y$. Thus $E(y) = \sigma y / (\epsilon h)$,
so that the free energy contribution from the region between the plates is 
 ${\cal E} =  \epsilon h \int_{-L}^{L} E^2(y) \, dy  = ( 2 \sigma^2 L^3 ) / ( 3\epsilon h).$
There is also a contribution to the free energy from the region outside the gap. 
The electric field $E(r)$ at distances $r \gg h$ satisfies the condition of flux conservation
 $\pi \epsilon  r E(r) = 2 L \sigma$ and the free energy 
 contribution from the region outside the gap is 
 $ {\cal E}^{\prime}  \sim \epsilon \int_{h}^{\alpha L} E^2(r) \pi r \, dr
  = 4 \sigma^2 L^2   \ln ( \alpha L / h )   / (\epsilon \pi )$,
 where we have replaced the inner and outer limits of integration by $h$ and $\alpha L$ respectively
 in order to prevent a divergent integral. This artifice accounts for the fact that the electric 
 field is not truly singular as $r \rightarrow 0$ but is determined by the pore geometry 
 in the region $r \sim h$. Moreover, in a real system, the $1/r$ decay of the 2D Coulomb field 
 must give way to a $1/r^2$ decay in 3D, once $r$ exceeds the supposed infinite
 extension of the slit in the $z$-direction. This dimension is of course 
 in practice finite and we denote it by $\alpha L$,
$\alpha$ being a large but finite aspect ratio.
 Therefore, in the limit $h/L \rightarrow 0$,
 ${\cal E}^{\prime} / {\cal E} \sim (h/L) \ln (L/h) \rightarrow 0$. 
 Thus, the force of interaction, $F$, in the limit $h \ll L$ 
 may be obtained from $2 F \; dh = - d {\cal E}$, which gives 
 $F \sim \sigma^2 L^3 /( 3 \epsilon h^2 )$ in agreement with Eq.~(\ref{totalF1}) when $\kappa L\ll 1$.

In the opposite extreme of Debye lengths short compared to $L$
(i.e. $h\ll \kappa^{-1}\ll L $)
Eq.~(\ref{totalF1}) implies that 
$F = 2 f_0 L / (\kappa h)^{2}   - 2 f_0 / (\kappa^3 h^2 )$.
 The first term represents the force in the absence of ion overspill
and, as expected, is proportional 
 to $L$. The second term represents a reduction of force 
 arising from overspill
 at each of the two edges which results in a drop in the ionic 
 contribution to the osmotic pressure. 
 This quantity is independent of the length $L$ 
 of the plates since field perturbations are confined to 
 distances $\sim \kappa^{-1}$ from the edges.
The reduction of the force corresponds 
 to a loss of osmotic pressure within a region of length $\kappa^{-1}$ 
 at each of the two plate edges. This consitutes an edge correction to the $1/h^2$ scaling law
of Philipse et al.\cite{philipse_algebraic_2013}, due to the finite size of the plates.
Our results for this edge correction can be used for plates of arbitrary shape as long as the radius
of curvature of the plate edges is large compared to the Debye length, so that a locally 2D analysis is appropriate.

\begin{figure}[t]     
  \centering
    \includegraphics[width=0.8\textwidth]{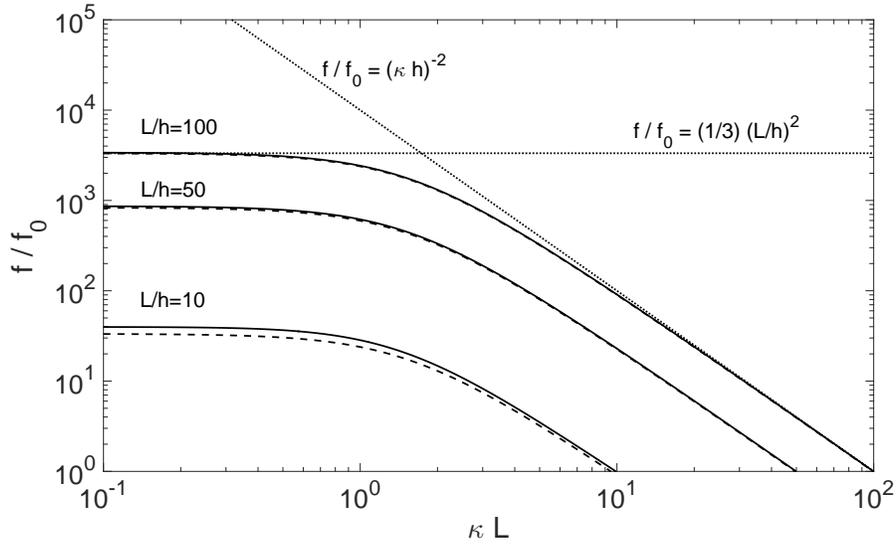}
      \caption{The normalized average disjoining pressure $f/f_0$ as a function of $\kappa L$ for fixed aspect ratios $L/h$, evaluated using Eq.~(\ref{totalF}) [solid lines] and Eq.~(\ref{totalF1}) [dashed lines].
      The dotted line of slope $-2$ corresponds to $f/f_0 = 1 /(\kappa h)^{2}$, the expected result in the absence of edge effects. The horizontal dotted line corresponds to $f/f_0 = (1/3) (L/h)^2$, the limiting value 
      at infinite Debye length. These limits are only shown for the case of $L/h = 100$ for clarity.}
      \label{fig_press}
\end{figure} 
Results for $f / f_0$  as a function of $\kappa L$, given by Eqs.~(\ref{totalF}) and (\ref{totalF1}),
are shown in Figure~\ref{fig_press}, for fixed
aspect ratio $L/h$.
This corresponds to a physical experiment in which the interaction force is measured 
for various electrolyte strengths. We see that at short Debye lengths ($\kappa L \gg 1$),
$f/f_0 = 1 / \kappa^2 h^2 = (1 / \kappa^2 L^2)(L/h)^2$ as reported 
by Philipse et al.\cite{philipse_algebraic_2013} 
and indicated by the dashed line in the figure.
However, as the Debye length becomes large compared to $L$, the average disjoining pressure does not diverge but approaches the limiting value $(f_0/3) (L/h)^{2}$, the expected result in the absence 
of mobile ions. 

The reduction in disjoining pressure due to edge effects should be readily observable in interactions between objects of small size, such as nanoparticles. Indeed, in measurements using the scanning force microscope (SFM)~\cite{todd_probing_2004} the force-distance graph was found to be well described by the Deryaguin approximation when the Debye length was less than $3$ nm but not so when the Debye length was increased to $9.6$ nm, a value larger than the $7$ nm apex radius of the SFM tip. The discrepancy increases with decreasing plate separation as one would expect from Eq.~(\ref{totalF}). The geometry 
in the experimental set up~\cite{todd_probing_2004} was axisymmetric, rather than plane 2D, making a quantitative comparison with Eq.~(\ref{totalF}) inappropriate. However, it is clear that the effects discussed in our analysis will also occur in the axisymmetric case.
\begin{acknowledgement}
J.D.S. thanks Ory Schnitzer for helpful discussions, and the Department of Applied Mathematics
\& Theoretical Physics, University of Cambridge, for hospitality.
\end{acknowledgement}
\providecommand{\latin}[1]{#1}
\providecommand*\mcitethebibliography{\thebibliography}
\csname @ifundefined\endcsname{endmcitethebibliography}
  {\let\endmcitethebibliography\endthebibliography}{}

\newpage
\begin{tocentry}
\begin{center}
\includegraphics[width=0.5\textwidth]{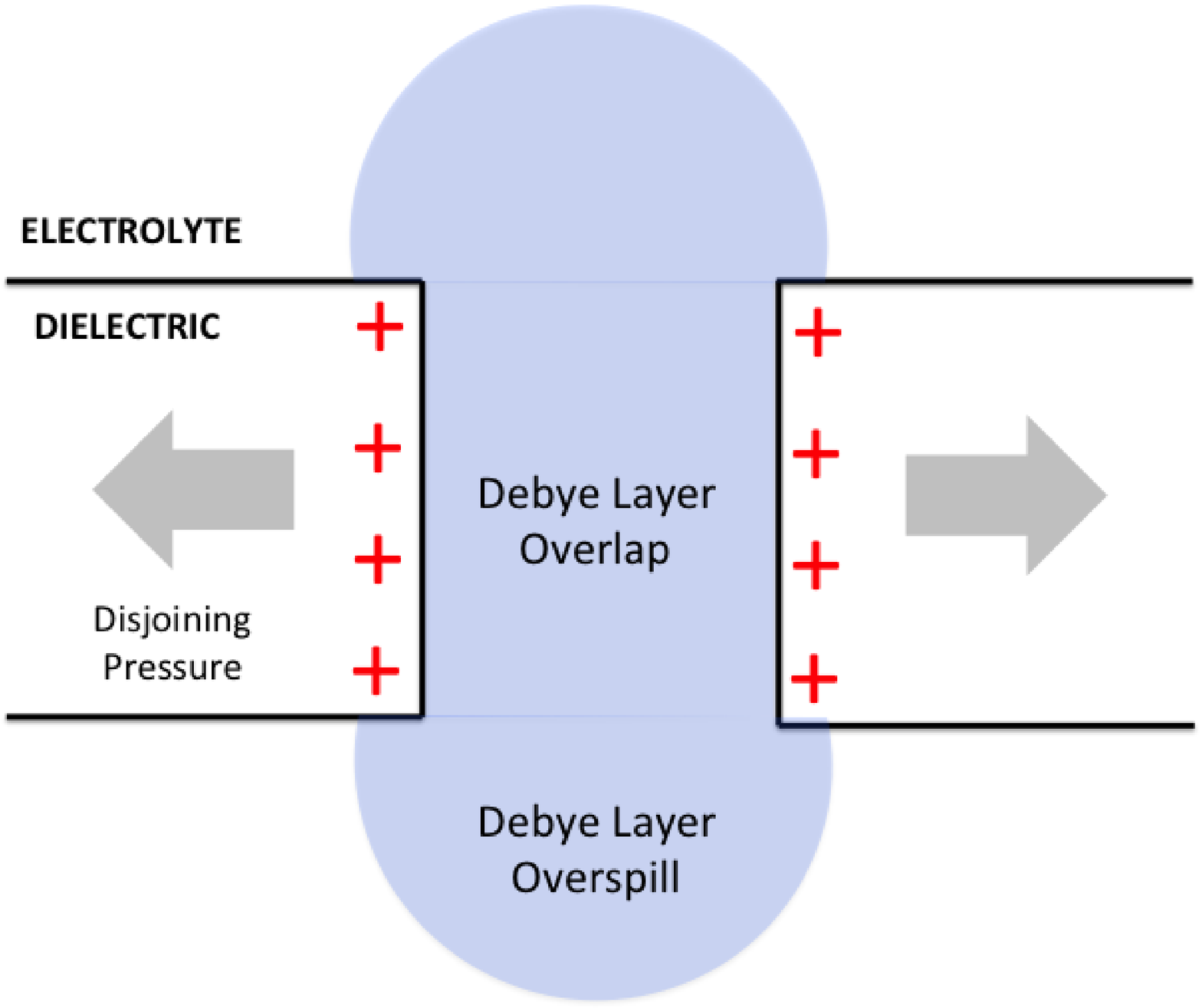}
\end{center}
\end{tocentry}
\end{document}